\documentclass[twocolumn,showpacs,preprintnumbers,amsmath,amssymb]{revtex4}

\usepackage{graphicx}
\usepackage{dcolumn}
\usepackage{bm}

\begin{document}

\preprint{APS/123-QED}

\title{Long-lived isomeric states in neutron-deficient thorium isotopes?}

\author{J. Lachner}
  \email{johannes.lachner@physik.tu-muenchen.de}
\author{I. Dillmann}
\altaffiliation{Excellence Cluster "Origin and Structure of the Universe", Technische Universit\"at M\"unchen, D-85748, Garching, Germany}
\author{T. Faestermann}
\author{G. Korschinek}
\author{M. Poutivtsev}
\altaffiliation{Excellence Cluster "Origin and Structure of the Universe", Technische Universit\"at M\"unchen, D-85748, Garching, Germany}
\author{G. Rugel}
 
\affiliation{Physik Department E12 and E15, Technische Universit\"at M\"unchen, D-85748 Garching, Germany}%

\date{\today}

\begin{abstract}
The discovery of naturally occurring long-lived isomeric states (t$_{1/2}$ $>$ 10$^8$ yr) in the neutron-deficient isotopes $^{211,213,217,218}$Th [Marinov et al., Phys. Rev. C \textbf{76}, 021303(R) (2007)] was re-examined using accelerator mass spectrometry (AMS). 
As AMS does not suffer from molecular isobaric background in the detection system, it is an extremely sensitive technique. 
In spite of our up to two orders of magnitude higher sensitivity we cannot confirm the discoveries of neutron-deficient thorium isotopes and provide upper limits for their abundances.

\end{abstract}
\maketitle

\section{\label{intro}Introduction}
Recently the discovery of "high spin super- and hyperdeformed" isomeric states in neutron-deficient thorium isotopes was claimed \cite{MRK07}. 
The authors investigated solutions of natural thorium and announced the discovery of long-lived isomers in $^{211,213,217,218}$Th. 
They conclude that the lower limit of the half lives must be 10$^8$~years and assume that these might belong to "high spin $K$-type isomers", which were "preferentially produced by heavy ion reactions".

These observations are extremely surprising for several reasons.
The ground states of the thorium isotopes in question are all $\alpha$ emitters with half lives ranging from 144\,ms
down to 117\,ns for the $N$=128 nucleus $^{218}$Th which has one of the shortest half lives ever 
measured for a nuclear ground state. 
In Table~\ref{tab:th} we list the available information on these thorium isotopes.
Even if the electromagnetic decay for the isomers claimed by \cite{MRK07} would be so slow due to $K$-hindrance, also the $\alpha$-decay would have to be slowed down due to angular momentum hindrance by at least 16 to 22 orders of magnitude to explain the long half lives required for their natural occurence.

\begin{table}[htb]
\caption{\label{tab:th}Decay properties of $^{211,213,217,218}$Th and known isomers, with excitation energy, isospin and parity.}
\renewcommand{\arraystretch}{1.2} 
\begin{ruledtabular}
\begin{tabular}{lcccl}
 Isotope & I$^\pi$ & E$_{ex}$ [keV] & t$_{1/2}$ & Ref. \\
 \hline
 $^{211}$Th 		& ? 					& 0			& (40 $^{+3}_{-1}$)\,ms 		& \cite{NDS211}\\
 $^{213}$Th 		& (5/2)$^-$ 	& 0			& (144 $\pm$21)\,ms 			& \cite{NDS213}\\
 $^{213}$Th$^m$ & (13/2)$^+$ & 1180 	& (1.4 $\pm$0.4)\,$\mu$s 	& \cite{KHH07}\\
 $^{217}$Th 		& (9/2$^+$)   & 0 	 	& (241 $\pm$5)\,$\mu$s 		& \cite{NDS217}\\
 $^{217}$Th$^m$ & (25/2$^+$)  & "2252+X"  & (67 $^{+17}_{-11}$)\,$\mu$s & \cite{KHA05}\\
 $^{218}$Th 		& 0$^+$ 			& 0			& (117 $\pm$9)\,ns 				& \cite{NDS218}\\
\end{tabular}
\end{ruledtabular}
\end{table}

Furthermore there is no way within the current understanding of nucleosynthesis how these states could have been produced before the formation of the solar system.

The stellar nucleosynthesis of isotopes heavier than iron occurs predominantly via neutron capture reactions in equal shares in the so-called slow and rapid processes ("$s$ process" and "$r$ process") \cite{bbfh57}. 
The neutron density in the $s$ process is in the range $N_n\approx$10$^{8}$cm$^{-3}$, forcing the reaction path to follow the line of $\beta$ stability. 
The $s$ process nucleosynthesis terminates in the reaction cycle $^{209}$Bi($n,\gamma$)$^{210}$Bi($\beta^-$)$^{210}$Po($\alpha$)$^{206}$Pb. 
In order to populate thorium isotopes in the $s$ process, a bridge of nuclear states with half lives in the order of years between $^{209}$Bi and these thorium isotopes would be necessary.

The long-lived primordial actinides $^{232}$Th, $^{235}$U and $^{238}$U are produced in the high neutron density scenario ($N_n\gg$10$^{20}$ cm$^{-3}$) of the $r$ process. 
The exact stellar production site is still under discussion, most probable scenarios are core-collapse supernovae or neutron star mergers. 
The starting seed for both, the $r$ and $s$ processes, are iron peak nuclei produced by previous fusion processes ("burning stages") in the star. 
Due to its large neutron density the $r$-process reaction path is driven far away from stability close to the region of the neutron drip-line. 

By this mechanism the $r$ process is able to climb within a few hundred milliseconds up to its termination point around $A\approx$260 ($Z$=94, Pu) where "fission recycling" by spontaneous and $\beta$-delayed fission sets in and transfers some material back into the mass region $A\approx$130. 
When the neutron production ends and/or the temperature decreases ("freeze-out") the short-lived radioactive nuclei decay back to stability via long $\beta$-decay chains. 
This produces the well-known observed solar $r$-abundance peaks at $A\approx$80, 130, and 195 \cite{ande89}, corresponding to the very neutron-rich $r$-process progenitor nuclei with $N$=50, 82, and 126.

Beyond $A$=210 these long decay chains will reach $\alpha$-emitters which are (almost) stable against $\beta$-decays. 
The transmutation of these radioactive isotopes will then proceed via $\alpha$-decay chains, ending either in bismuth and the lead isotopes or in the long-lived actinides $^{232}$Th, $^{234}$U, $^{235}$U and $^{238}$U. 
In any case the nuclides finally populated in the $r$ process have a considerably higher $A/Z$ ratio than the values $A/Z\le$218/90 for the thorium isotopes in question.

Thus the findings of Ref.\cite{MRK07} motivated us to repeat the measurements with an independent method.

\section{Experimental technique}
Accelerator mass spectrometry (AMS) is a highly sensitive method to determine small traces of radionuclides without interfering background from molecules.
For the measurement we used our AMS-setup at the Maier-Leibnitz-Laboratory in Garching.

\begin{figure}[h]
\includegraphics[width=0.48\textwidth]{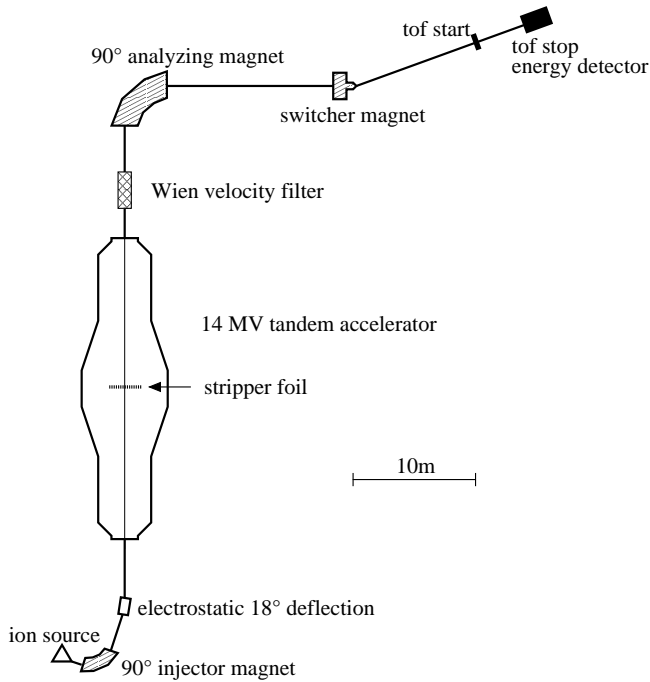}
\caption{AMS setup for the detection of $^{211,213,217,218}$Th.}
\end{figure}

The thorium sample was pressed in a graphite cathode for the sputter ion source from where ThC$^-$-molecules of different masses were extracted. 
The sample was prepared by mixing ThO$_2$ (Merck LotNr.\,2353854) with graphite powder.

The first mass selection of the molecules is performed with the 90$^{\circ}$ injector magnet, which has a mass resolution of 1/200. 
For an additional energy selection on the low-energy-side we use a 18$^{\circ}$ electrostatic deflection.

The terminal voltage of the MP tandem accelerator was set at voltages around 11\,MV and thus the negatively charged molecules are accelerated to an energy of about 11\,MeV. 
The thin carbon stripper foil (4\,$\mu$g/cm$^2$) in the high voltage terminal destroys these molecules completely. 
Positive ions in high charge states are produced and accelerated further on the high-energy side. The thorium ions had typically an energy of around 120\,MeV. 
The beam transport system was adjusted with $^{209}$Bi$^{10+}$ ions from $^{209}$BiO$^-$. 
This setting was verified with $^{232}$Th$^{10+}$ from $^{232}$ThC$^-$. 

On the high-energy side of the accelerator the charge state 10$^+$ was selected by a Wien velocity filter and the high resolution 90$^{\circ}$ analyzing magnet. 
In order to transport different ions, all the magnetic elements after the accelerator were untouched and only the terminal voltage and the voltage of the Wien-filter were adjusted. 
The parameters of the beam transport for the neutron-deficient thorium isotopes were calculated with the masses given in \cite{MRK07}.
The acceptance of our beam transport system has a value of about $\Delta m/m\approx \pm$4$\cdot$10$^{-4}$.

The ions are identified by a time-of-flight spectrometer, where the start signal is delivered by a microchannelplate detector.
An ionization chamber determines the energy loss of the ions and a silicon surface barrier counter measures the residual energy and acts as the stop detector simultaneously.
The setup of the detection system is described in more detail in \cite{WFG00}.

From time to time a beam of $^{232}$Th, attenuated by a factor of 10$^6$ with two low transmission grids, is sent to the detector for normalization of the flux of the thorium isotopes. 
The count rate of the unattenuated beam of $^{232}$Th corresponds to about 10$^8$\,Hz for the various measurements. 
The number of detected events of the less abundant thorium isotopes is normalized to this $^{232}$Th particle current. 
Scattering inside the attenuator grids and thus loss of transmission was taken into account by the measurement of the sum of the daughter products $^{228}$Th and $^{228}$Ra. 
They cannot be distinguished by the described detection method. 
As RaC$^{-}$ and ThC$^{-}$ have a different formation rate in a cesium sputter ion source, the measured $A$=228 particle flux is corrected according to the data in \cite{ZNG94,tims04}. 
The employed ThO$_2$ is more than 15 years old, so $^{228}$Ra and $^{228}$Th should be in equilibrium. 
Thus the transmission of the different thorium isotopes of interest versus the attenuated $^{232}$Th is accounted for by multiplying the $^{232}$Th yield with a factor of 1.2 to 1.4. 
The runtime for each isotope lasted several hours.

\section{Results}

None of the four neutron-deficient thorium isotopes could be detected with an abundance of 1 to 10$\cdot$10$^{-11}$ as suggested in \cite{MRK07}. 
 
For the two nuclides $^{211}$Th and $^{217}$Th there were no events in the expected regions of the time-of-flight signal and the energy signal. 
Background events are detected at a wide range of different time-of-flights and energies due to scattered ions for $^{232}$Th in the 11$^+$ or 10$^+$ charge state. 
One event was measured in the expected range for $^{213}$Th and $^{218}$Th respectively, which would correspond to an abundance of these isotopes on the level of several 10$^{-13}$. 
But these single events can not be distinguished from the outspread background.

Upper limits for the yield were estimated from the measurement values of run time, particle flux and transmission, taking into account 1\,$\sigma$ errors of these quantities as a consequence of several uncertainties. 
Furthermore statistical analysis of small signals according to \cite{fc98} was included. 

The value for the additional extraction of $^{228}$Ra has an uncertainty of 10\% \cite{ZNG94}. 
The uncertainty of the attenuation is assumed with 5\%. 
The particle currents of mass 228 and mass 232, used to determine the transmission and the flux of the thorium ions, are subject to a statistical fluctuation of 5\%. 

Our measurements contradict the observations of Marinov et al.\cite{MRK07}.
The resulting upper limits (see Table~\ref{upperlimit}) are about one order of magnitude below the values given in \cite{MRK07} and are in accordance with results from Dellinger et al.\cite{DFG08}.

\begin{table}[htb]
\caption{\label{upperlimit}AMS-results for the abundance of $^{211,213,217,218}$Th, upper limits given at the 68.3\% confidence level.}
\renewcommand{\arraystretch}{1.2} 
\begin{ruledtabular}
\begin{tabular}{ccc}
 Isotope & events & upper limit for $^{A}$Th/$^{232}$Th \\
 \hline
 $^{211}$Th 		& 0	& 9.6$\cdot$10$^{-13}$	\\
 $^{213}$Th 		& 1	& 1.2$\cdot$10$^{-12}$	\\
 $^{217}$Th 		& 0 	& 6.6$\cdot$10$^{-13}$	\\
  $^{218}$Th 		& 1	& 2.4$\cdot$10$^{-12}$	\\
\end{tabular}
\end{ruledtabular}
\end{table}

\section{Discussion}

The values of our AMS measurements in thorium samples demonstrate that in nature no neutron-deficient isotopes of thorium at masses 211, 213, 217 and 218 exist with abundances of 1 to 10$\cdot$10$^{-11}$ relative to $^{232}$Th.
Measurements were stopped as soon as it became clear that the result of the concentration of neutron-deficient thorium isotopes in our sample is about an order of magnitude below the values from \cite{MRK07}. 
So our sensitivity is limited by the run time more than by the background because of scattered $^{232}$Th ions.

There is also no scenario of nucleosynthesis besides the $r$ process which is able to produce isotopes beyond $^{209}$Bi in such amounts that they would be somehow detectable on earth. 
However, the $r$ process cannot reach neutron-deficient thorium isotopes like $^{211,213,217,218}$Th in any way since they are located in the midst of pure $\alpha$ emitters. 
The lightest thorium isotope that can be reached by the $r$-process $\beta$-decay chains is $^{226}$Th.

The argument to attribute these long-lived isomers to $K$ isomers \cite{WD99} can also be disproved. 
Most $K$ isomers were identified around $A\approx$180 (e.g. $^{180}$Ta$^m$) where large mid-shell deformations with neutrons and protons occupying high-lying orbitals can be found. 
The heaviest $K$ isomers found up to now are in $^{250}$Fm \cite{GEE73}, $^{256}$Fm \cite{HGH89}, $^{252}$No \cite{SHH07}, $^{254}$No \cite{GEE73,TKS06}, and $^{270}$Ds \cite{HHA01}. 
These isomeric states have half lives in the order of $\mu$s to ms. 
Prolate deformations are an important condition for the existence of $K$ isomers. 
This requirement is not fulfilled for the neutron-deficient thorium isotopes. 
The aforementioned superheavy nuclei exhibit ground-state quadrupole deformations between $\beta_2$=0.221 and 0.246 (prolate), whereas for $^{211}$Th the maximum deformation is calculated to be $\beta_2$=-0.105 (oblate deformation) \cite{MNM95}. 
Also $^{213}$Th and $^{217}$Th are estimated to be slightly oblate, whereas the even-even nucleus $^{218}$Th -- two neutrons away from the shell closure at $N$=126 -- is expected almost spherical ($\beta_2$=0.008). 
Thus no $K$ isomers should exist in $^{211,213,217,218}$Th.

\begin{acknowledgments}
This research was supported by the DFG cluster of excellence "Origin and Structure of the Universe". 
We thank A. T\"urler (Institut f\"ur Radiochemie, TU M\"unchen) for providing us with suitable sample material. 
\end{acknowledgments}

\end{document}